\documentclass[jap,preprint,groupedaddress,showpacs]{revtex4}
\usepackage{multirow}
\usepackage[fleqn]{amsmath}
\usepackage{graphicx}
\usepackage{docs}         
\usepackage{bm}
\usepackage{color}
\usepackage[abs]{overpic}
\usepackage[colorlinks=true,linkcolor=blue,citecolor=blue,urlcolor=black]{hyperref}%
\expandafter\ifx\csname package@font\endcsname\relax\else
 \expandafter\expandafter
 \expandafter\usepackage
 \expandafter\expandafter
 \expandafter{\csname package@font\endcsname}%
\fi

\usepackage{setspace}
\begin{document}
\setstretch{1.5}
\title{Ballistic Phonon Transport in Ultra-Thin Silicon Layers:\\ 
Effects of Confinement and Orientation}
\author{Hossein Karamitaheri, Neophytos Neophytou, and Hans Kosina}
\affiliation{Institute for Microelectronics,
  Technische Universit\"at Wien, Gu{\ss}hausstra{\ss}e 27--29/E360,
  A-1040 Wien, Austria\\
  \hspace*{0.02\linewidth}
  $\mathrm{E-mail:}\{\mathrm{karami}~|~\mathrm{neophytou}~|~\mathrm{kosina}\}\mathrm{@iue.tuwien.ac.at}$ \hspace{0.02\linewidth}}

\date{\today}

\begin{abstract}
We investigate the effect of confinement and orientation on the phonon transport properties of ultra-thin silicon layers of thicknesses between $1~\mathrm{nm}-16~\mathrm{nm}$. We employ the modified valence force field method to model the lattice dynamics and the ballistic Landauer transport formalism to calculate the thermal conductance. We consider the major thin layer surface orientations $\{100\}$, $\{110\}$, $\{111\}$, and $\{112\}$. For every surface orientation, we study thermal conductance as a function of the transport direction within the corresponding surface plane. We find that the ballistic thermal conductance in the thin layers is anisotropic, with the $\{110\}/\textless110\textgreater$ channels exhibiting the highest and the $\{112\}/\textless111\textgreater$ channels the lowest thermal conductance with a ratio of about two. We find that in the case of the $\{110\}$ and $\{112\}$ surfaces, different transport orientations can result in $\sim 50\%$ anisotropy in thermal conductance. The thermal conductance of different transport orientations in the $\{100\}$ and $\{111\}$ layers, on the other hand, is mostly isotropic. These observations are invariant under different temperatures and layer thicknesses. We show that this behavior originates from the differences in the phonon group velocities, whereas the phonon density of states is very similar for all the thin layers examined. We finally show how the phonon velocities can be understood from the phonon spectrum of each channel. Our findings could be useful in the design of the thermal properties of ultra-thin Si layers for thermoelectric and thermal management applications. 
\end{abstract}

\maketitle

\section{Introduction}

The thermal conductivity of Si is dominated by phonon transport and has a relatively high value of $\kappa_l=148~\mathrm{W/mK}$. Such high conductivity is beneficial for some applications such as heat management in electronic devices~\cite{Goodson95}, but unwanted for other applications such as thermoelectricity. Low dimensional Si materials, such as nanowires, ultra-thin layers, and nanoporous Si, on the other hand, have demonstrated record low thermal conductivities of $\kappa_l=1-2~\mathrm{W/mK}$, reaching the amorphous limit~\cite{Boukai08,Hochbaum08,Chen01,Chen08b,Wu02,Tang10,Yu10,Hopkins10,Song04}. The thermal conductivity in Si is carried by phonons of a few nanometers to a few micrometers in wavelength~\cite{Zebarjadi12}, and boundary scattering is very effective in suppressing the propagation of low frequency (long wavelength) phonons in Si~\cite{Ponomareva07,Yang08,Wang09b,Liangraksa11,Oh12,Martin09,Liu05,Liu06}.  

Although the two order of magnitude reduction in the thermal conductivity is attributed to boundary scattering, an additional reduction can be achieved from changes in the phonon mode structure due to geometrical confinement. Indeed, the phonon mode dispersion undergoes strong modifications in nanostructures~\cite{Lazarenkova02,Broido04,Duchemin11,Ramayya08}. The thermal conductivity in bulk Si is isotropic, however, in low-dimensional materials the choice of geometrical features such as surface orientation, transport orientation, and confinement length scale (i.e. thickness or diameter) can result in different phonon modes. These differences in the phonon modes affect the phonon group velocities and the scattering processes, and introduce variations in the thermal conductance~\cite{Aksamija10,Turney10}. The proper choice of structure geometries can, therefore, lead to different thermal properties and can allow design optimization for the applications of interest. 

Very few studies on the effect of geometrical features such as the surface orientation, transport orientation, and layer thickness on the thermal conductivity of ultra-thin-body layers (UTBs), however, can be found in the literature. These mostly focus on layers of larger thicknesses of 10s of nanometers, or employ bulk Si phonon dispersions, whose validity could be debatable for layers with thicknesses down to a few nanometers. Aksamija et al. have theoretically discussed the effects of confinement and orientation of thin Si membranes using the bulk phonon dispersion and Boltzmann transport theory~\cite{Aksamija10}. That work elucidated the importance of geometry, and indicated that there can be indeed a factor of two difference in the thermal conductivity once a proper channel is chosen (the $\{110\}$ versus the $\{100\}$ surfaces in that case). In order to properly understand how modifications in the phonon mode structure will affect the thermal transport of ultra-thin layers in the sub-ten $~\mathrm{nm}$ thickness scale, however, a model that goes beyond the bulk dispersion, and properly captures the effect of confinement on the phonon modes is required. The importance of the complete phonon dispersion details in addressing thermal transport in nanostructures has been stressed in several publications~\cite{Mingo03,Tian11}. Experimental data could only be explained once these details were taken into consideration~\cite{Mingo03b}.
    
In this work, we employ the modified valence force field (MVFF) method~\cite{Sui93} to address the effects of structural confinement and transport orientations on the phonon dispersion, group velocity, and ballistic thermal conductance of Si thin layers of thickness from $1~\mathrm{nm}$ to $16~\mathrm{nm}$. For a complete study we investigate various surface orientations and transport orientations. We consider the $\{100\}$, $\{110\}$, $\{111\}$ and $\{112\}$ surface orientations, and for each of these surfaces we calculate thermal conductance as a function of the transport orientation. We find that the variation in the thermal conductance between channels of different geometries can be up to a factor of two. This is true for the choice of different surfaces, but also different transport orientations within the same surface. This observation is only weakly dependent upon the layer thickness. The $\{110\}/\textless110\textgreater$ channel exhibits the highest and the $\{112\}/\textless111\textgreater$ channel the lowest thermal conductance, almost $\sim 50\%$ lower. We further show that any variations observed are a consequence of the phonon group velocities which are anisotropic, whereas the density of phonon modes does not show strong anisotropy. We provide explanations for the group velocity behavior through features of the phonon modes. 

The paper is organized as follows: i) In section~\ref{s:Approach} we describe the MVFF method for the calculation of the phonon bandstrucutre, and the Landauer method for phonon transport calculations. In section~\ref{s:Results} we present the results, and in section~\ref{s:Analysis} we provide explanations and discussions. Finally in section~\ref{s:Conclusions} we conclude.

\section{Approach}
\label{s:Approach}
For bulk studies, the most frequent model traditionally employed for phonon dispersion calculations is the Debye model, in which the phonon dispersion is described by three acoustic branches, one longitudinal, and two transverse modes. More sophisticated models that could describe the full phonon dispersions of bulk as well as nanostructures is the valence force field method (the Keating model~\cite{Keating66}), the Tersoff inter-atomic potential model~\cite{Tersoff89}, the adiabatic bond charge model~\cite{Hepplestone11}, as well as first principle calculations. In this work, for the calculation of the phononic bandstructure we employ the modified valence force field method~\cite{Sui93}, which is an extension of the Keating model. In this method the interatomic potential is modeled by the following bond deformations: bond stretching, bond bending, cross bond stretching, cross bond bending stretching, and coplanar bond bending interactions~\cite{Sui93}. The model accurately captures the bulk Si phonon spectrum as well as the effects of confinement~\cite{Paul10}. 

In the MVFF method, the total potential energy of the system is defined as~\cite{Paul10}:
\begin{equation}
U\approx \frac{1}{2}\sum_{i\in N_{\mathrm{A}}} \left[ \sum_{j\in nn_i} U_{\mathrm{bs}}^{ij} + \sum_{j,k\in nn_i}^{j\neq k} \left (U_{\mathrm{bb}}^{jik}+U_{\mathrm{bs-bs}}^{jik}+U_{\mathrm{bs-bb}}^{jik}\right) +\sum_{j,k,l\in COP_i}^{j\neq k\neq l} U_{\mathrm{bb-bb}}^{jikl} \right]
\label{e:MVFFPotential}
\end{equation}
where $N_{\mathrm{A}}$, $nn_i$, and $COP_i$ are the number of atoms in the system, the number of the nearest neighbors of a specific atom $i$, and the coplanar atom groups for atom $i$, respectively. $U_{\mathrm{bs}}$, $U_{\mathrm{bb}}$, $U_{\mathrm{bs-bs}}$, $U_{\mathrm{bs-bb}}$, and $U_{\mathrm{bb-bb}}$ are the bond stretching, bond bending, cross bond stretching, cross bond bending stretching, and coplanar bond bending interactions, respectively. The terms $U_{\mathrm{bs-bs}}$, $U_{\mathrm{bs-bb}}$, and $U_{\mathrm{bb-bb}}$ are an addition to the usual Keating model~\cite{Keating66}, which can only capture the Si phononic bandstructure in a limited part of the Brillouin zone. As indicated in Ref.~\cite{Paul10} the introduction of these additional terms provides a more accurate description of the entire Brillouin zone.  

In this formalism we assume that the total potential energy is zero when all the atoms are located in their equilibrium positions. Under the harmonic approximation, the motion of atoms can be described by a dynamic matrix as~\cite{Karamitaheri12}:
\begin{equation}
D=\left[ D_{3\times 3}^{ij} \right]= \left[ \frac{1}{\sqrt{M_iM_j}}\times \left \{ \begin{array}{lll} D_{ij} & {} & ,i\neq j \\ {} & {} & {} \\-\displaystyle \sum _{l\neq i}D_{il} & {} & ,i=j \end{array} \right. \right]
\end{equation}
where dynamic matrix component between atoms $i$ and $j$ is given by~\cite{Paul10}:
\begin{equation}
D_{ij}=\left[ \begin{array}{ccc} D_{xx}^{ij} & D_{xy}^{ij} & D_{xz}^{ij} \\ D_{yz}^{ij} & D_{yy}^{ij} & D_{yz}^{ij} \\ D_{zx}^{ij} & D_{zy}^{ij} & D_{zz}^{ij} \end{array} \right]
\end{equation}
and
\begin{equation}
D_{mn}^{ij}=\frac{\partial^2 U_{\mathrm{elastic}}}{\partial r_m^i \partial_n^j},~~~~~~ i,j\in N_{\mathrm{A}}~\mathrm{and}~m,n\in [x,y,z]
\end{equation}
is the second derivative of the potential energy with respect to the displacement of atoms $i$ and $j$ along the $m$-axis and the $n$-axis, respectively. $U_{\mathrm{elastic}}$ is
the potential that associated with the motion of only two atoms $i$ and $j$, whereas the
other atoms are considered frozen (unlike $U$, which is the potential when all atoms are
allowed to move out of their equilibrium position). To compute this: 1) We start with $U$
from Eq.~\ref{e:MVFFPotential}. 2) We fix the positions of all atoms except atoms $i$ and $j$. 3) We compute the inter-atomic potential due to all bond deformations that result from interaction between both of these two atoms, and sum them up to obtain $U_{\mathrm{elastic}}$. All other inter-atomic potential terms that result from interactions due to atom $i$ alone, or atom $j$ alone, are not considered, since all double derivatives taken with respect to $\partial^2/\partial r_m^i \partial_n^j$, give zero.

After setting up the dynamic matrix, the following eigenvalue problem is solved for the calculation of the phononic dispersion:
\begin{equation}
D+\sum_l D_l~\exp{\left({i{\bf \overrightarrow{q}}.\Delta {\bf \overrightarrow{R}}_l}\right )}-\omega ^2({q})I=0
\label{e:PhEigen}
\end{equation}
where $D_l$ is the dynamic matrix representing the interaction between the unit cell and its neighboring unit cells separated by ${\Delta \bf \overrightarrow{R}}_l$~\cite{Karamitaheri12}. Using the phononic dispersion, the phonon density of states (DOS) and the ballistic transmission (number of modes at given energy) are calculated by~\cite{Datta05Book}:
\begin{equation}
DOS(\omega)=\sum_{\alpha}DOS_{\alpha}(\omega)=\sum_{\alpha}\sum_{q}\delta\left( \omega - \omega_{\alpha}(q)\right)
\end{equation}
and
\begin{equation}
\overline{T}_{\mathrm{ph}}(\omega)=M(\omega)=\frac{h}{2}\sum_{\alpha,q}\delta \left( \omega - \omega_{\alpha}(q) \right) v_{g,\alpha}(q)\Big\vert_{\parallel}
\end{equation}
where $v_{g,\alpha}(q)\Big\vert_{\parallel}$ is the parallel component of the group velocity $V_{g,\alpha}(q)=\frac{\partial \omega_{\alpha}(q)}{\partial q}$ along the transport orientation. In the expressions above, at a specific frequency $\omega$ the sum runs over all phonon modes ($\alpha$) and all phonon momenta ($q$) of the two-dimensional momentum space.

Once the transmission is obtained, the ballistic lattice thermal conductance is calculated within the framework of the Landauer theory as~\cite{Landauer57,Jeong12}:
\begin{equation}
\kappa_l=\frac{1}{h}\int_{0}^{+\infty}\overline{T}_\mathrm{ph}(\omega)\hbar\omega\left(\frac{\partial n(\omega)}{\partial T}\right)\ d(\hbar\omega)
\label{e:Conductance}
\end{equation}
where $n(\omega)=(\mathrm{e}^{\hbar \omega/k_\mathrm{B}T}-1)^{-1}$ is the Bose-Einstein distribution function. Alternatively, the energy integral in Eq.~\ref{e:Conductance} can be transformed into a summation over $q$-space where the thermal conductance is evaluated as:
\begin{equation}
\kappa_l=\sum_{\alpha,q}\kappa_{l,\alpha}(q)
\label{e:Conductance2}
\end{equation}
where the $q$- and $\alpha$-dependent thermal conductance is defined as:
\begin{equation}
\kappa_{l,\alpha}(q)=\frac{1}{2}\frac{2\pi}{\Delta q_{\parallel}}\frac{2\pi}{\Delta q_{\perp}} v_{g,\alpha}(q)\Big\vert_{\parallel} \hbar \omega_{\alpha}(q) \frac{\partial n\left( \omega_{\alpha}(q) \right)}{\partial T}
\end{equation}
In this work, both Eq.~\ref{e:Conductance} and Eq.~\ref{e:Conductance2} are employed depending on whether we compute $\omega$- or $q$-dependent data.

We note that in this work we calculate the ballistic thermal \emph{conductance} of the thin layers, not the \emph{conductivity} which assumes diffusion of phonons after undergoing all relevant scattering mechanisms. Our intention in this work is to specifically investigate the influence of the confined phonon bandstructure on the anisotropy of the phonon transport.	
\section{Results}
\label{s:Results}
\begin{figure}[htb]
\begin{center}
 \includegraphics[width=0.95\linewidth]{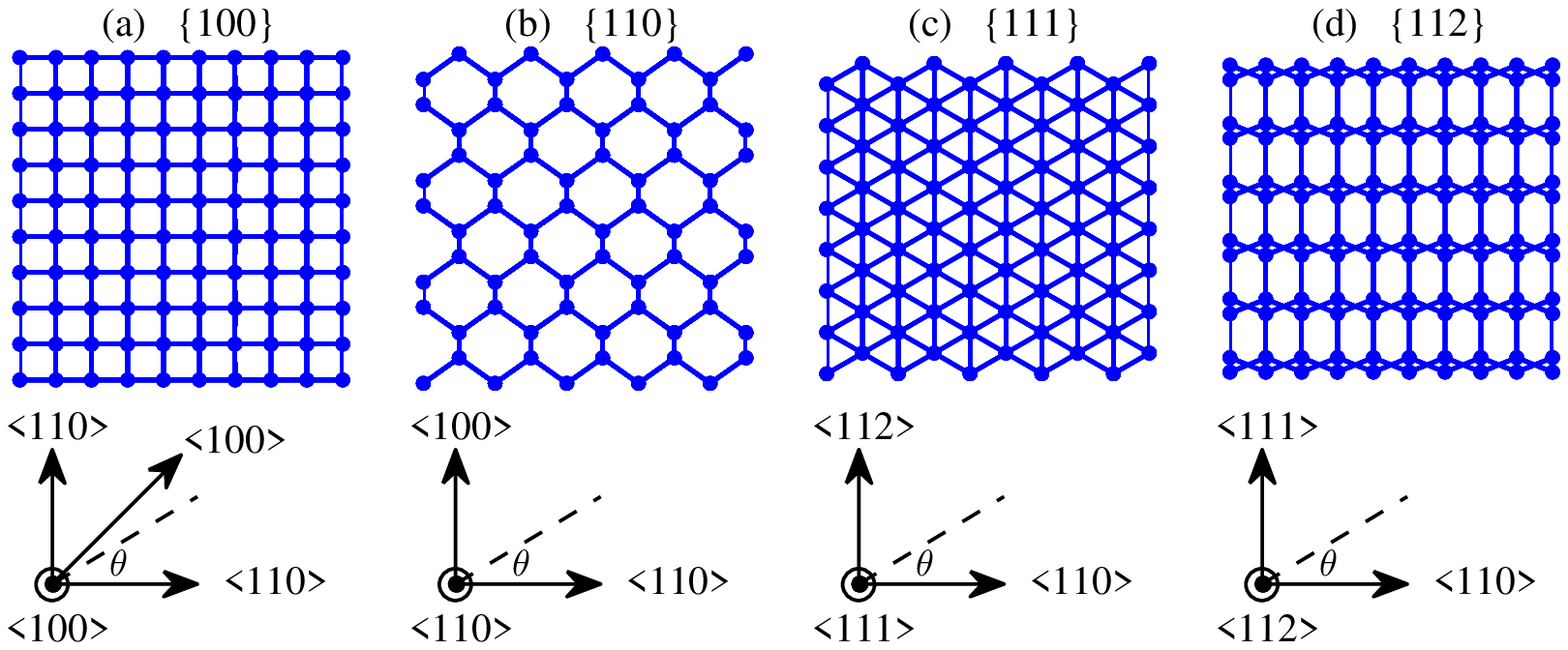}
\caption{The atomistic structure in the cross sections of the different thin layers investigated (a) $\{100\}$, (b) $\{110\}$, (c) $\{111\}$, and (d) $\{112\}$ surface. In all cases the $x$-axis is along the $\textless110\textgreater$ transport direction. We consider different transport orientations by varying the angle $\theta$ between 0 and $\pi$.}
\label{f:Figure1}
\end{center}
\end{figure}

Figure~\ref{f:Figure1} shows the geometrical cross sections of the thin layers considered. These are the $\{100\}$, $\{110\}$, $\{111\}$, and $\{112\}$ surface orientations. In all cases, we consider the $x$-axis to be the $\textless110\textgreater$ orientation, and define the angle $\theta$ of the transport direction counter-clockwise from the $x$-axis. Below we present a complete analysis by calculating the phononic properties and thermal conductance as a function of the angle $\theta$ for all the surface orientations mentioned. We also vary the layer thickness $H$ from $1~\mathrm{nm}$ to $16~\mathrm{nm}$. We calculate the phononic dispersion, density of states, ballistic transmission, and effective group velocity of the different structures.  

Figure~\ref{f:Figure2} shows the transmission functions for the four layer surface orientations of interest along two particular transport orientations for each case, that, as we will show below, provide the lowest and the highest thermal conductance for that particular surface. The layer thickness in all cases is $2~\mathrm{nm}$. In the case of the thin layer with $\{100\}$ surface orientation in Fig.\ref{f:Figure2}-a, we consider the $\{100\}/\textless110\textgreater$ and the $\{100\}/\textless100\textgreater$ transport channels. The transmissions of the two channels are almost the same, indicating negligible anisotropy. In the case of the thin layer with $\{111\}$ surface orientation in Fig.~\ref{f:Figure2}-c, we consider the $\{111\}/\textless110\textgreater$ and the $\{111\}/\textless112\textgreater$ transport channels. Again in this case, the transmissions are almost the same. 

The transmission function of the thin layers with $\{110\}$ and $\{112\}$ surfaces, on the other hand, is orientation dependent. For the $\{110\}$ surface thin layers in Fig.~\ref{f:Figure2}-b, the $\{110\}/\textless110\textgreater$ channel (blue line) shows the highest transmission function, and the $\{110\}/\textless100\textgreater$ channel (red-dotted line) the lowest. An even larger difference is observed in the case of the $\{112\}$ surface thin layers in Fig.~\ref{f:Figure2}-d. The highest transmission is observed for the $\{112\}/\textless110\textgreater$ channel (blue line), and the lowest for the $\{112\}/\textless111\textgreater$ channel (red-dotted line). The difference in the transmission of the channels in different transport orientations is largest for energies between $10-30~\mathrm{meV}$ for both, the $\{110\}$ and the $\{112\}$ thin layers.

Using the transmission functions extracted from the bandstructures, we calculate the ballistic lattice thermal conductance using the Landauer formula for the thin layers with the four different surface orientations of interest. We calculate the thermal conductance as a function of the transport orientation by varying the angle $\theta$ from 0 to $\pi$. The thermal conductances for all cases shown in Fig.~\ref{f:Figure3} are calculated for room temperature. We calculate the conductance of thin layers for thicknesses of 1, 2, 4, 8 and $16~\mathrm{nm}$. With symbols we denote the high symmetry orientations using the Miller index notation, i.e. $\textless110\textgreater$ - circle, $\textless111\textgreater$ - star, $\textless112\textgreater$ - triangle, and $\textless100\textgreater$ - square. We mark these orientations on the $16~\mathrm{nm}$ thin layer result in Fig.~\ref{f:Figure3}. In all cases, the conductance increases linearly as the thickness increases because the thicker layers contain more phonon modes that contribute to the thermal conductance. With regards to anisotropy, for the thin layers with $\{100\}$ surface in Fig.~\ref{f:Figure3}-a, the conductance has a maximum along the $\textless100\textgreater$ direction (square), and a minimum is along the $\textless110\textgreater$ direction (circle), although the difference is small (only $\sim 5\%$). Interestingly, this observation is the same for all thicknesses considered. The conductance of the channels with $\{110\}$ surface is shown in Fig.~\ref{f:Figure3}-b. The conductance is highest in the $\textless110\textgreater$ transport orientation ($\theta=0$, circle), and is lowest for the $\textless100\textgreater$ channels ($\theta=\pi/2$, square). The variation between the maximum and minimum, however, in this case is $\sim 30\%$ for the $1~\mathrm{nm}$ thin layer, and decreases to $20\%$ for the $16~\mathrm{nm}$ layer. The conductance of channels with $\{111\}$ surface is shown in Fig.~\ref{f:Figure3}-c. The conductance in this case also peaks along the $\textless110\textgreater$ direction (circle) and it is lowest along the $\textless112\textgreater$ direction (triangle). The variation of the conductance with transport orientation in this case is negligible for the thinner layers, but increases to $\sim10\%$ in the $16~\mathrm{nm}$ case. The thermal conductance for channels with $\{112\}$ surface is shown in Fig.~\ref{f:Figure3}-d. The maximum and minimum conductance is observed along $\textless110\textgreater$ (circle) and $\textless111\textgreater$ (star), respectively. Channels with this surface indicate the largest variation in thermal conductance compared to other surfaces. The difference varies from $\sim 40\%$ for the $1~\mathrm{nm}$ layers to $\sim 30\%$ for the $16~\mathrm{nm}$ layers. Overall, considering all surfaces and transport orientations, the maximum thermal conductance is observed for the $\{110\}/\textless110\textgreater$ channels, and the minimum for the $\{112\}/\textless111\textgreater$ channels. Interestingly, however, regardless of surface orientation, the thermal conductance is high in $\textless110\textgreater$ direction. This agrees well with previous works on silicon nanowires, where it is reported that the $\textless110\textgreater$ oriented nanowires have the highest thermal conductance~\cite{Markussen08,Paul11,Karamitaheri13}. A similar conclusion was found for thin layers of larger sizes~\cite{Aksamija10}. As we shall explain below, the phonon dispersions along the $\textless110\textgreater$ orientations are more dispersive compared to other orientations, which yield higher group velocities and, therefore, highest thermal conductance.
 
Figure~\ref{f:Figure4} shows the thermal conductance of the $H=2~\mathrm{nm}$ layers as a function of temperature. For every surface orientation we show two transport orientations, the one with the maximum and the one with the minimum conductance (as in Fig.~\ref{f:Figure2}). The conductance increases with temperature as expected from a ballistic quantity, and starts to saturate around $300~\mathrm{K}$. The reason is that the thermal conductance in Eq.~\ref{e:Conductance} can be also expressed as: 
\begin{equation}
\kappa_l=\frac{k_{\mathrm{B}}^2T\pi^2}{3h} \int_{0}^{+\infty}\overline{T}_\mathrm{ph}(\omega) W_{\mathrm{ph}}(\hbar \omega) d(\hbar \omega)
\end{equation}
where 
\begin{equation}
W_{\mathrm{ph}}=\frac{3}{\pi^2}\left( \frac{\hbar \omega}{k_{\mathrm{B}}T} \right)^2 \frac{\partial n}{\partial (\hbar \omega)}
\end{equation}
is the so-called phononic window function~\cite{Jeong12}. The phonon energy spectrum of Si extends up to $\sim 65~\mathrm{meV}$, and for sufficiently high temperatures the phononic window function is nearly constant within the entire $\sim 65~\mathrm{meV}$ energy range, as also shown in Ref.~\cite{Jeong12}. This causes the thermal conductance to saturate. Figure~\ref{f:Figure4} shows that the $\{110\}/\textless110\textgreater$ channel has the largest conductance, and the $\{112\}/\textless111\textgreater$ channel the smallest in the entire temperature range. The conductances of the other channels lie in between and do not deviate significantly from one another. The same trend is observed for the $H=16~\mathrm{nm}$ channels (inset of Fig.~\ref{f:Figure4}), although the spread is smaller. Below, we provide explanations for this geometry dependence in terms of the phonon bandstructure, by extracting the phonon density of states and the effective group velocity.
\section{Analysis}
\label{s:Analysis}

The ballistic thermal conductance in the Landauer formalism is determined by the product of the density of states and the group velocity. In Fig.~\ref{f:Figure5} we plot the density of states for thin layers of thickness $H=2~\mathrm{nm}$ and the four different surface orientations of interest. Although some differences are observed for the different surface orientations, especially in the low frequency range, the overall values and trends are very similar. The inset of Fig.~\ref{f:Figure5} shows the density of states for layers of thickness $H=16~\mathrm{nm}$. In this case a much smaller variation is observed as expected, since the phonon density of states depends at first order on the number of atoms, and layers of the same thickness contain a similar amount of atoms. At smaller thicknesses the different arrangement of atoms can result in slightly different numbers of atoms for different surfaces, but as the thickness increases the crystal becomes more uniform and any variations are eliminated. In general, of course, the arrangement of atoms, the coupling between them, and the type of interactions they have can also influence their density of states. But as we show in Fig.~\ref{f:Figure5}, such effects
are only important on the density of states at very thin sizes, i.e. $H=2~\mathrm{nm}$,
and even then, they are small. We note that also in the case of Si nanowires, our previous
work has demonstrated a similar result, namely that even for nanowires with cross section
sizes down to $H=6~\mathrm{nm}$, the density of states is orientation independent~\cite{Karamitaheri13}. From this we conclude that the variation in the thermal conductance and transmission does not originate from the difference in the density of states. 

\begin{figure}[htb]
\begin{center}
 \includegraphics[width=0.70\linewidth]{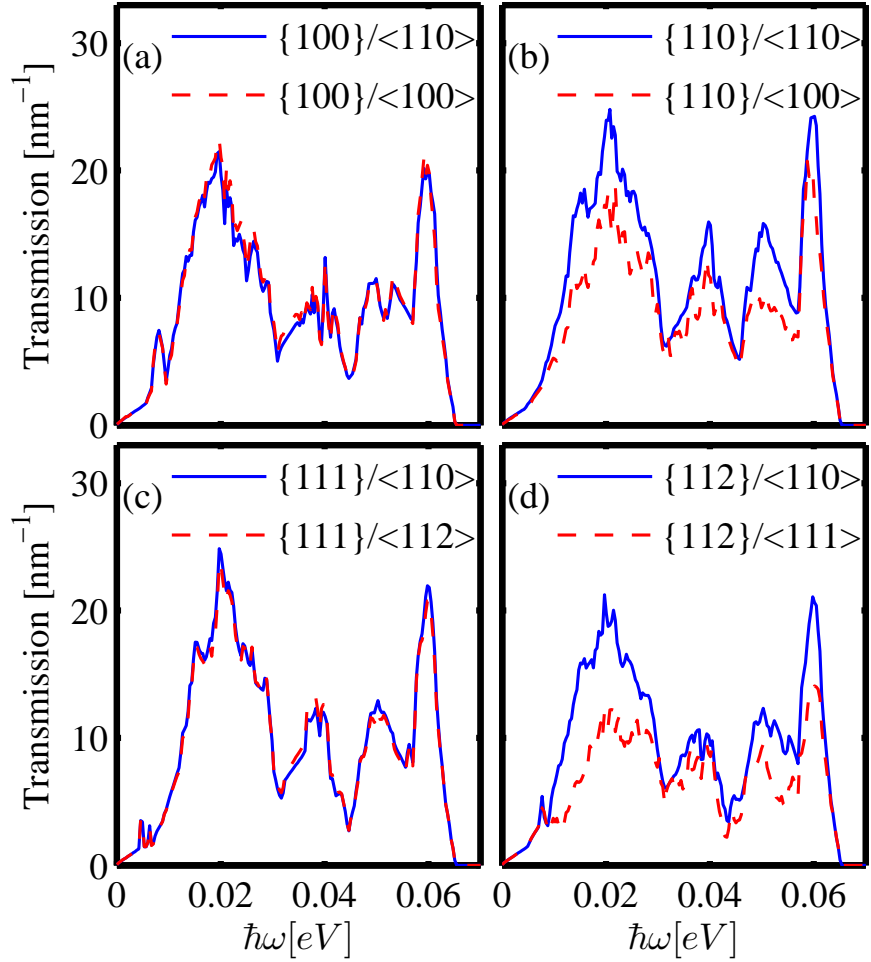}
\caption{Transmission function versus energy for thin layers of thickness $H=2~\mathrm{nm}$ with (a) $\{100\}$, (b) $\{110\}$, (c) $\{111\}$, and (d) $\{112\}$ surfaces, for two transport orientations in each case. The different transport orientations are the ones that yield the highest (blue-solid) and the lowest (red-dashed) thermal conductance in the corresponding surface orientation.}
\label{f:Figure2}
\end{center}
\end{figure}

\begin{figure}[htb]
\begin{center}
 \includegraphics[width=0.95\linewidth]{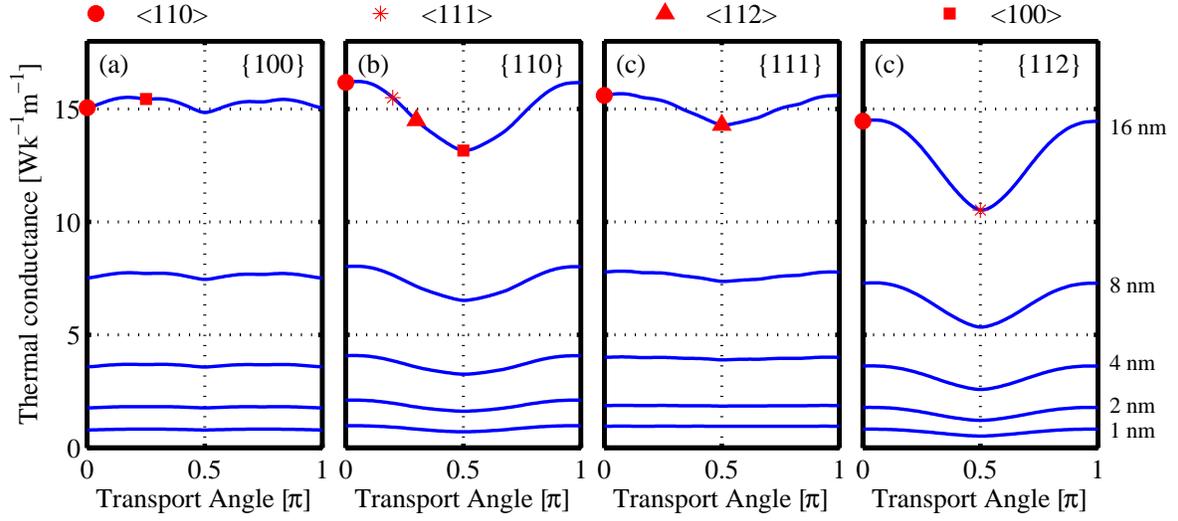}
\caption{Ballistic lattice thermal conductance for different thin layers with (a) $\{100\}$, (b) $\{110\}$, (c) $\{111\}$, and (d) $\{112\}$ surfaces. The angle $\theta$ as shown in Fig.~\ref{f:Figure1} specifies the transport orientation. Some of the high symmetry orientations are denoted by symbols. Results for different layers thicknesses are shown. From bottom to top, the thicknesses are 1, 2, 4, 8, and $16~\mathrm{nm}$.}
\label{f:Figure3}
\end{center}
\end{figure}

\begin{figure}[htb]
\begin{center}
 \includegraphics[width=0.70\linewidth]{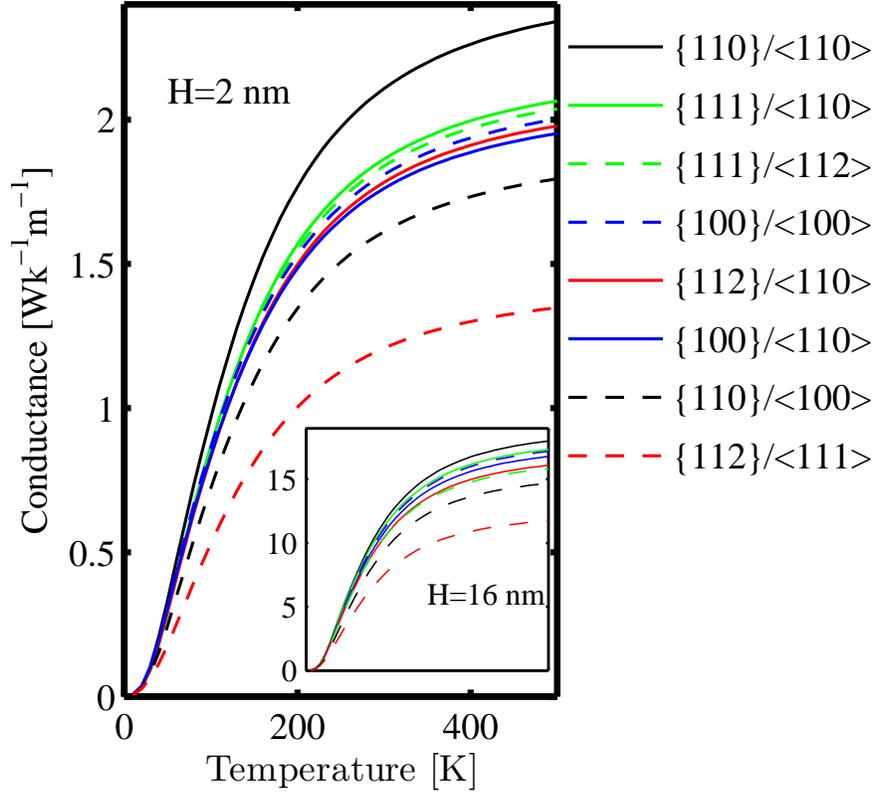}
\caption{The thermal conductance of thin layers of $2~\mathrm{nm}$ thickness for various surface and transport orientations, as a function of temperature. The transport orientations are the ones that result in the highest (solid) and lowest (dashed) thermal conductance for the respective surface. Inset: The same quantity for thin layers of $H=16~\mathrm{nm}$.}
\label{f:Figure4}
\end{center}
\end{figure}

\begin{figure}[htb]
\begin{center}
 \includegraphics[width=0.70\linewidth]{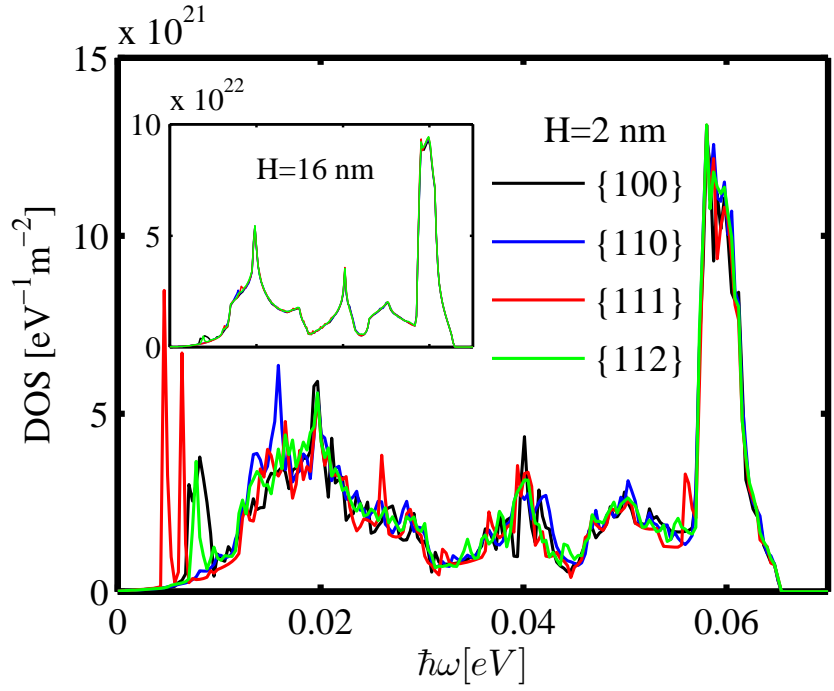}
\caption{The density of phonon states for thin layers of thickness $H=2~\mathrm{nm}$ for different surface orientations. Inset: The density of phonon states for $H=16~\mathrm{nm}$.}
\label{f:Figure5}
\end{center}
\end{figure}

\begin{figure}[htb]
\begin{center}
 \includegraphics[width=0.70\linewidth]{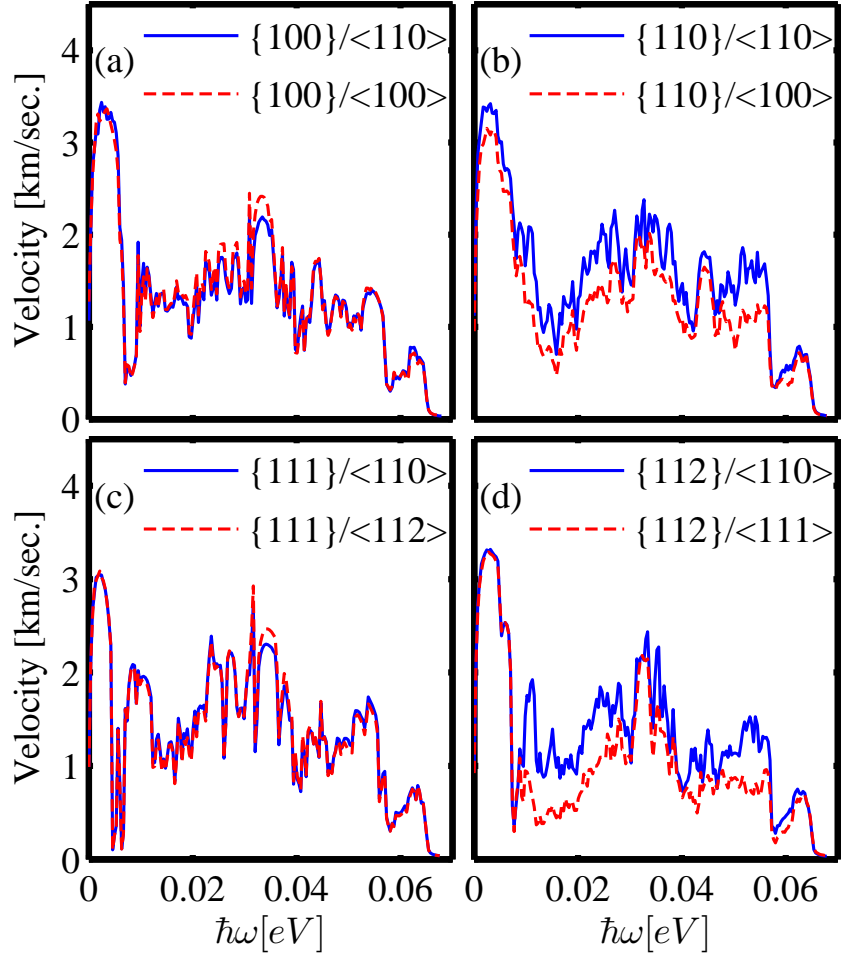}
\caption{The effective group velocity versus energy for thin layers of thickness $H=2~\mathrm{nm}$ for (a) $\{100\}$, (b) $\{110\}$, (c) $\{111\}$, and (d) $\{112\}$ surfaces, for two transport orientations in each case. The transport orientations chosen are the ones that result in the highest (blue-solid) and the lowest (red-dashed) thermal conductance for the given surface orientation.}
\label{f:Figure6}
\end{center}
\end{figure}

In Fig.~\ref{f:Figure6} we plot the second quantity that influences the transmission and conductance, which is related to the velocity of the phonon states. We define the effective group velocity at a specific energy $E=\hbar \omega$ as the weighted average of the velocities of the phonon states, with the weighting factor being the density of states: 
\begin{equation}
\ll V_g(\omega)\gg=\frac{\displaystyle{\sum_{\alpha,q}}v_{g,\alpha}(q)\Big\vert_{\parallel} \delta\left( \omega -\omega_{\alpha}(q)\right)}{\displaystyle{\sum_{\alpha,q}}\delta\left( \omega -\omega_{\alpha}(q)\right)}
\label{e:Veff}
\end{equation}
The velocity of a phonon is in general a function of the subband index, the frequency, and the wavenumber $q$. The quantity in Eq.~\ref{e:Veff} averages over the subband index and the wavenumber and thus provides a quantity that depends only on frequency (or energy). Similar "effective" quantities have also been used in thermal conductivity calculations in different works as well~\cite{Zou01,Mingo03,Jeong10}. However, in our actual calculations we utilize all the information of the phonon spectrum. This quantity is orientation-dependent, in contrast to the density of states, and indicates how dispersive the modes are. The velocity  is calculated along the transport direction. The density of states times the effective group velocity is proportional to the transmission function. Therefore, the differences in the transmission functions should be seen in the effective group velocities of the channels, since the density of states is the same for all channels of the same thickness. Figure~\ref{f:Figure6} shows the effective group velocities of the channels considered. Figures~\ref{f:Figure6}-a and~\ref{f:Figure6}-c show the effective group velocities of thin layers with $\{100\}$ and $\{111\}$ surfaces along the two different orientations with the lower and highest thermal conductance for each surface. The two different cases for each surface are almost identical, as in the case of the transmission functions in Fig.~\ref{f:Figure2}-a and~\ref{f:Figure2}-c. Figures~\ref{f:Figure6}-b and~\ref{f:Figure6}-d show the effective group velocities for channels with $\{110\}$ and $\{112\}$ surfaces, respectively. A variation is observed for the different channels, which causes the difference in the transmission functions shown earlier in Fig.~\ref{f:Figure2}-b and~\ref{f:Figure2}-d. 

\begin{figure}[htb]
\begin{center}
 \includegraphics[width=0.66\linewidth]{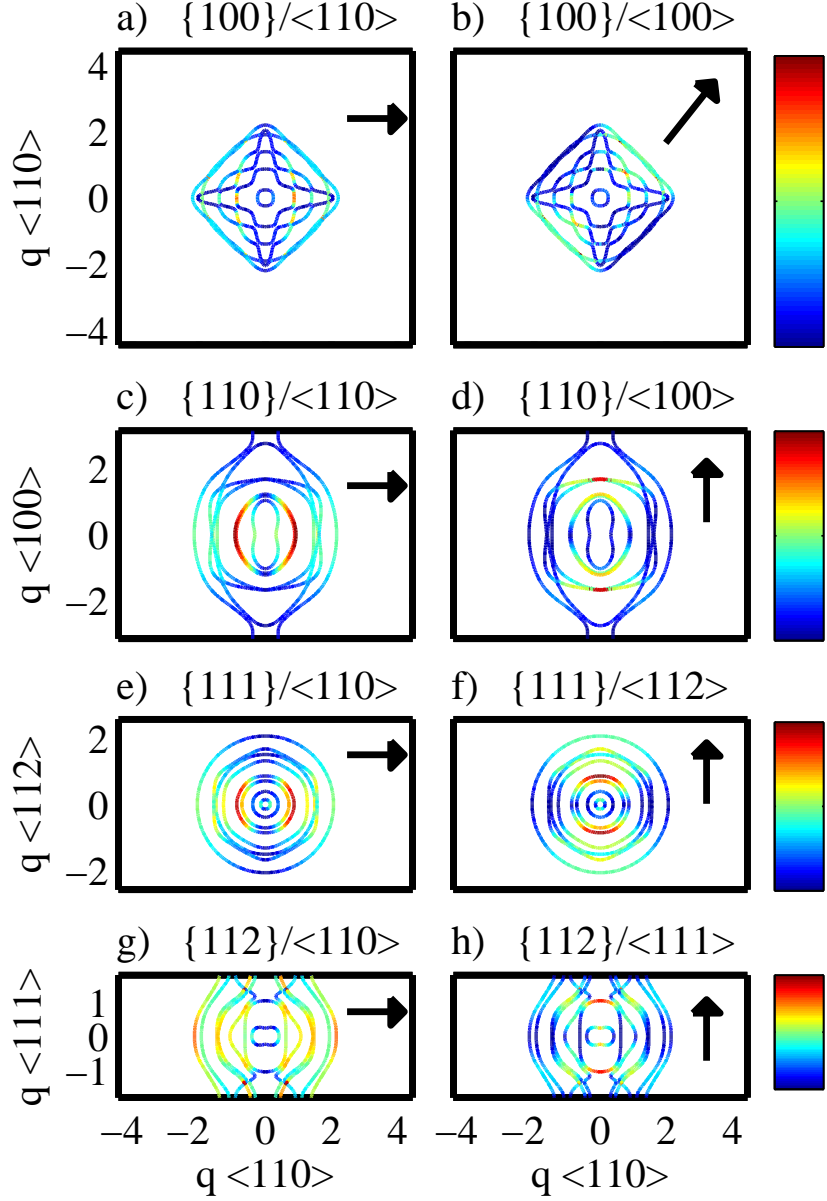}
\caption{Energy contours at $E=10~\mathrm{meV}$ for thin layers of different surface and transport orientations. (a, b) $\{100\}$ surface, (c, d) $\{110\}$ surface, (e, f) $\{111\}$ surface, (g, h) $\{112\}$ surface. The transport orientations chosen are the ones that result in the highest (left) and the lowest (right) thermal conductance for the given surface. The color indicates the contribution of each $q$-state to the thermal conductance at $E=10~\mathrm{meV}$ at $300~\mathrm{K}$ along the different transport orientations in each thin layer (indicated by the arrow). The red color shows the highest value and the blue the lowest.}
\label{f:Figure7}
\end{center}
\end{figure}

The anisotropy (or isotropy) of the effective group velocity originates from the phonon bandstructure. In Fig.~\ref{f:Figure7} we show contour plots of the phonon bandstructure at $E=\hbar \omega=10~\mathrm{meV}$ for all eight channels considered in Fig.~\ref{f:Figure2} and Fig.~\ref{f:Figure6} for layer thickness $H=2~\mathrm{nm}$. This is an energy value at which the most significant differences for the channels with $\{110\}$ and $\{112\}$ surfaces appear. It turns out that what we present for this energy is a good indicator of the anisotropic behavior of the entire energy spectrum, most of which contributes to thermal conductance at room temperature. The lines represent the different modes at that energy, whereas the colormap indicates the contribution of each $q$-state to the ballistic thermal conductance at room temperature in the transport orientation of the specific channel of interest, as indicated by the arrow in each case. Elongation of contour lines along a specific direction provides high phonon group velocities in the perpendicular direction, and consequently high thermal conductance. This is very similar to the low effective mass and high velocities of carriers in an ellipsoidal band along the direction of the short axis in the case of electronic transport. Figures~\ref{f:Figure7}-a and~\ref{f:Figure7}-b show the energy contours for the $\{100\}$ surface in the $\textless110\textgreater$ and $\textless100\textgreater$ transport orientations, respectively (indicated by the arrow). Despite the square shape of the contour, which indicates that there is different symmetry in the two orientations of interest, the contours are elongated similarly in both directions, which results in a similar thermal conductance for both channels. This is the case for almost the entire energy spectrum (although at higher energies we have many more modes and more complex contour shapes). In the case of the $\{111\}$ surface in Fig.~\ref{f:Figure7}-e and~\ref{f:Figure7}-f, a highly symmetric contour provides very similar transmission functions and thermal conductances along the $\textless110\textgreater$ and $\textless112\textgreater$ directions. The largest differences in the thermal conductance are observed for thin layers with $\{110\}$ and $\{112\}$ surfaces in Fig.~\ref{f:Figure7}-c,~\ref{f:Figure7}-d and Fig.~\ref{f:Figure7}-g,~\ref{f:Figure7}-h, respectively. In both cases, the contours at energy $E=10~\mathrm{meV}$ are clearly elongated along the vertical axis. This results in a larger phonon group velocity along the horizontal axis, and finally a higher thermal conductance, as also indicated by the colormap. This is especially evident for the $\{112\}$ surface, where the contour of the $\textless110\textgreater$ channel in Fig.~\ref{f:Figure7}-g is colored much closer to red (higher conductance value) than the $\textless111\textgreater$ channel in Fig.~\ref{f:Figure7}-h, indicating much larger phonon group velocities. This causes the thermal transmission and conductance of the $\{112\}/\textless110\textgreater$ channel to be higher than that of the $\{112\}/\textless111\textgreater$ channel shown in Fig.~\ref{f:Figure2}-d and Fig.~\ref{f:Figure3}-d.      

\begin{figure}[htb]
\begin{center}
 \includegraphics[width=0.70\linewidth]{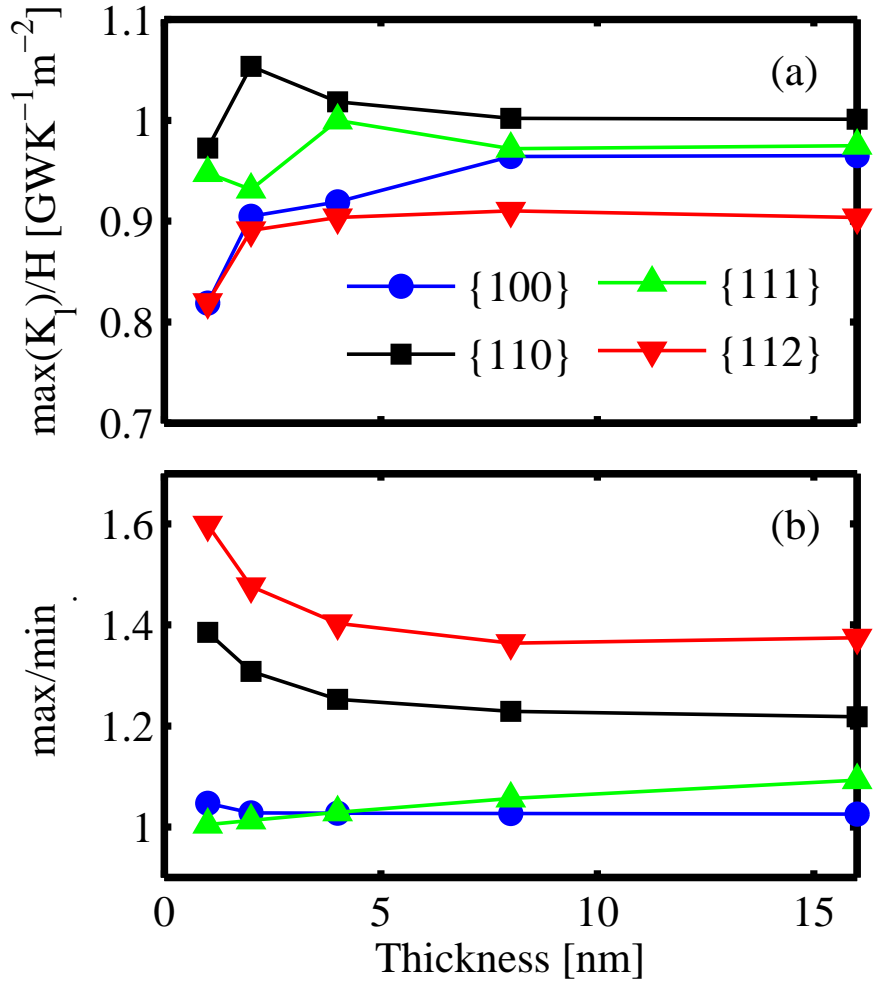}
\caption{(a) The maximum value of the conductance in thin layers of different surface orientations at $300~\mathrm{K}$ normalized by the thickness, versus layer thickness. (b) The ratio of the maximum to the minimum thermal conductance for different surface orientations, versus their thickness at $300~\mathrm{K}$. For each surface the transport orientations with the maximum and minimum conductance values are chosen.}
\label{f:Figure8}
\end{center}
\end{figure}

Figure~\ref{f:Figure7} explains the origin of anisotropy in the ballistic thermal conductance of thin layers with $2~\mathrm{nm}$ thickness. We point out, however, that such effects also hold for all the thicknesses we examine, e.g. up to $H=16~\mathrm{nm}$. The anisotropic behavior depends weakly on the layer thickness. The ballistic conductance increases linearly as the layer thickness increases due to the increased number of atoms which results in a larger number of phonon modes, but the anisotropy does not change significantly. This is illustrated in Fig.~\ref{f:Figure8}-a which shows the ballistic thermal conductance for each of the four surface orientations examined, normalized by the thickness of the layer. For each surface we only consider the direction showing the maximum conductance. We observe that in all cases the normalized conductance is constant, even down to a thickness of $H\sim 5~\mathrm{nm}$. Below $H\sim 5~\mathrm{nm}$, variations of the order of $\sim 10-20\%$ are observed for all channels. From this, it follows that other than the reduction in the size of the phonon spectrum with thickness scaling, no significant changes in the shape of the phonon structure are observed, at least not significant to introduce changes in the thermal conductance. This is also supported by Fig.~\ref{f:Figure8}-b, which depicts the ratio of the maximum to the minimum thermal conductance that can be achieved for each surface. Similarly to Fig.~\ref{f:Figure8}-a, the anisotropy does not change with layer thickness even down to $H\sim 5~\mathrm{nm}$. Again, below $H\sim 5~\mathrm{nm}$, differences of the order of $10-20\%$ can be observed.  

\begin{figure}[htb]
\begin{center}
 \includegraphics[width=0.70\linewidth]{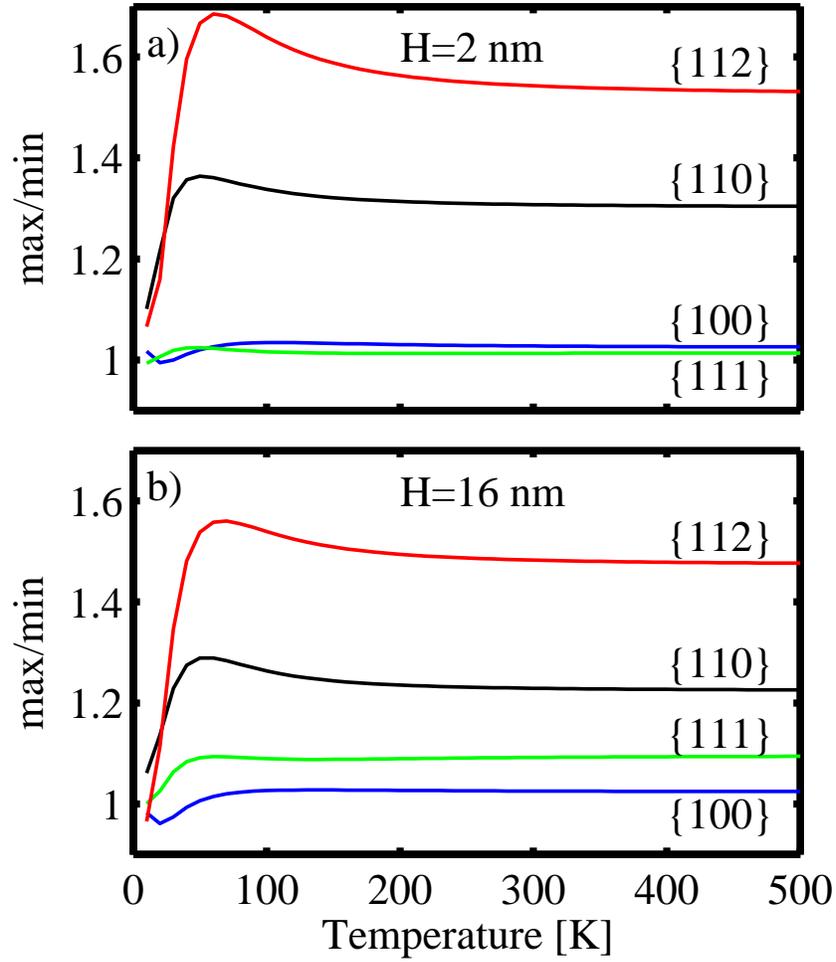}
\caption{The ratio of maximum to minimum thermal conductance of thin layers for different surfaces versus temperature. For each surface the transport orientation with the maximum and minimum conductance values are chosen. Thicknesses are (a) $H=2~\mathrm{nm}$ and (b) $H=16~\mathrm{nm}$.}
\label{f:Figure9}
\end{center}
\end{figure}

This anisotropy observed is not only a function of thickness, but also of temperature. Figure~\ref{f:Figure9} shows the ratio of the maximum to the minimum ballistic thermal conductance for the four surface orientations of interest, again by choosing the appropriate transport orientations. Figure~\ref{f:Figure9}-a and~\ref{f:Figure9}-b show results for $H=2~\mathrm{nm}$ and $H=16~\mathrm{nm}$, respectively. The maximum anisotropy (up to $60\%$) is observed for the $\{112\}$ surface, followed by the $\{110\}$ surface (up to $30\%$), whereas the $\{111\}$ and $\{100\}$ surfaces are more or less isotropic (the ratio stays $\sim 1$). This holds for most of the temperature range we examine, even down to $100~\mathrm{K}$. Below $100~\mathrm{K}$, the ratio approaches unity in all cases, because at this temperature the main contribution to thermal conductance comes from the acoustic branches at low energy, which are more isotropic. This is clearly observed in the low energy range of the transmissions in Fig.~\ref{f:Figure2}, in which the thermal conductivity is isotropic.

Finally, we need to mention that this work focused on the influence of bandstructure on the anisotropic behavior of the thermal transport properties of ultra-thin Si layers. Thus, we employed accurate phonon bandstructures, but utilized a rather simplified ballistic transport formalism, which ignores the effects of phonon scattering. Our intent is to provide a qualitative indication of the anisotropic behavior of phonon transport in thin layers. Employing atomistic phonon bandstructures and a fully diffusive transport formalism that accounts of the energy, momentum, and bandstructure dependence of each scattering event would be computationally very expensive, and will be the topic of subsequent studies. Our results, however, point out that a factor of two variation in phonon transport can be achieved once the channel geometry is optimized. These findings agree qualitatively well with diffusive phonon transport calculations that indicate the superiority of the thermal conductivity of the $\{110\}/\textless110\textgreater$ channel over other geometries, and the low thermal conductance for the $\{111\}/\textless110\textgreater$ and $\{112\}/\textless111\textgreater$ channels~\cite{Aksamija10}. They also agree with calculations for Si NWs, which indicate the
beneficial $\textless 110\textgreater$ transport orientation to heat transport, compared to other orientations~\cite{Markussen08,Paul11,Karamitaheri13}. When it comes to comparing to experimental results, however,unfortunately we could not identify any works in the literature that perform systematic thermal conductivity measurements in such ultra-thin layers ($H<16~\mathrm{nm}$) and in various confinement and transport orientations. Most experimental works on thermal conductivity consider relatively thick layers of thicknesses in the order of 10s-100s of nanometers and primarily on $\{100\}$ layers. In thicker layers the phonon modes are almost bulk-like and one cannot observe the anisotropic phonon confinement effects that lead to bandstructure modifications and conductance variations. In addition, the influence of various scattering mechanisms make the thermal conductivity of thicker layers more isotropic, and hide the results of bandstructure anisotropy (that ballistic simulations fully capture).

Our findings, however, are useful in understanding phonon transport in ultra-thin Si layers, and with regards to applications, could provide guidance in either maximizing heat transport as in the case of thermal management, or minimizing heat transport as in the case of thermoelectrics. For example, for electronic applications, we mention that for $p$-type nanoelectronic channels, transport in the $\{110\}/\textless110\textgreater$ orientation is beneficial compared to other orientations~\cite{NeoAPL11,NeoNL09}. This is also the case for the power factor in the case of thermoelectric devices~\cite{NeoJAP12}. In the former case, however, for electronic devices large thermal conductivity is necessary in order to remove the heat from the device, otherwise the mobility is degraded. The large thermal conductivity of the $\{110\}/\textless110\textgreater$ channel, therefore, could be advantageous for $p$-type electronic devices. In the latter case, for thermoelectric devices channels with low thermal conductivity are needed in order to reduce losses and increase thermoelectric efficiency. The large thermal conductivity of the $\{110\}/\textless110\textgreater$ channel, therefore, could counteract the benefit of its larger power factor, and this channel might not be the optimal for thermoelectric $p$-type Si devices.
\section{Conclusions}
\label{s:Conclusions}
The ballistic thermal conductance and its dependence on surface and transport orientations in ultra-thin silicon layers from $1~\mathrm{nm}$ to $16~\mathrm{nm}$ in thickness is investigated using the modified valence force field method and the Landauer formalism. The ballistic conductance of thin layers with $\{100\}$ and $\{111\}$ surface orientations is almost isotropic for all transport orientations. An anisotropy in the transport orientation of the order of $60\%$ and $40\%$ is observed for $\{112\}$ and $\{110\}$ channels respectively, due to the asymmetry in their phonon mode structure. In terms of absolute values, the $\{110\}/\textless110\textgreater$ channel has the highest thermal conductance, and the $\{112\}/\textless111\textgreater$ channel the lowest (almost $50\%$ lower). Interestingly, for all surfaces, the $\textless110\textgreater$ transport orientation shows the highest conductance. We finally show that these observations are layer thickness-independent, as well as temperature-independent. This anisotropy of transport for each surface is observed for temperatures above $100~\mathrm{K}$, whereas for lower temperatures the anisotropy is reduced. Our results can be useful in understanding the contribution of the phonon dispersion in the thermal conductivity of ultra-thin Si layers, as well as in the design of efficient thermal management and thermoelectric devices.    
\section*{Acknowledgments}
This work was supported by the European Commission, grant 263306 (NanoHiTEC).

\newpage
\clearpage{\textbf{Figure 1}}
\vspace*{3cm}

\newpage
\clearpage{\textbf{Figure 2}}
\vspace*{3cm}

\newpage
\clearpage{\textbf{Figure 3}}
\vspace*{3cm}

\newpage
\clearpage{\textbf{Figure 4}}
\vspace*{3cm}

\newpage
\clearpage{\textbf{Figure 5}}
\vspace*{3cm}

\newpage
\clearpage{\textbf{Figure 6}}
\vspace*{3cm}

\newpage
\clearpage{\textbf{Figure 7}}
\vspace*{0.01cm}

\newpage
\clearpage{\textbf{Figure 8}}
\vspace*{3cm}

\newpage
\clearpage{\textbf{Figure 9}}
\vspace*{3cm}

\end{document}